\documentclass[australian,showpacs,showkeys,superscriptaddress]{revtex4-1}
\usepackage[T1]{fontenc}
\usepackage[latin9]{inputenc}
\usepackage{amsmath}
\usepackage{graphicx}
\usepackage{babel}
\begin{document}

\title{Third-order transport coefficients for localised and delocalised
charged-particle transport}

\author{Peter W. \surname{Stokes}}
\email[Electronic address: ]{peter.stokes@my.jcu.edu.au}

\affiliation{College of Science and Engineering, James Cook University, Townsville,
QLD 4811, Australia}

\author{Ilija \surname{Simonovi\'{c}}}

\affiliation{Institute of Physics, University of Belgrade, PO Box 68, 11080 Zemun,
Belgrade, Serbia}

\author{Bronson \surname{Philippa}}

\affiliation{College of Science and Engineering, James Cook University, Cairns,
QLD 4870, Australia}

\author{Daniel Cocks}

\affiliation{College of Science and Engineering, James Cook University, Townsville,
QLD 4811, Australia}

\author{Saša \surname{Dujko}}

\affiliation{Institute of Physics, University of Belgrade, PO Box 68, 11080 Zemun,
Belgrade, Serbia}

\author{Ronald D. \surname{White}}

\affiliation{College of Science and Engineering, James Cook University, Townsville,
QLD 4811, Australia}
\begin{abstract}
We derive third order transport coefficients of skewness for a phase-space
kinetic model that considers the processes of scattering collisions,
trapping, detrapping and recombination losses. The resulting expression
for the skewness tensor provides an extension to Fick's law which
is in turn applied to yield a corresponding generalised advection-diffusion-skewness
equation. A physical interpretation of trap-induced skewness is presented
and used to describe an observed negative skewness due to traps. A
relationship between skewness, diffusion, mobility and temperature
is formed by analogy with Einstein's relation. Fractional transport
is explored and its effects on the flux transport coefficients are
also outlined.
\end{abstract}
\maketitle

\section{\label{sec:Introduction}Introduction}

Very little data regarding third order transport coefficients (the
skewness tensor) can be found in the literature. This is understandable,
since they have not been included in the interpretations of traditional
swarm experiments. There is, however, a growing interest regarding
these transport coefficients, partially due to estimations that third
order transport coefficients could be measured in the present or near
future \citep{Pitchford1990,Vrhovac1999}. It is also considered that
third order transport coefficients would be very useful, in combination
with transport coefficients of a lower order, for determination of
cross section sets, by means of inverse swarm procedure \citep{Pitchford1990,Vrhovac1999}.
Third order transport coefficients are also required for the conversion
of the hydrodynamic transport coefficients into transport data measured
in steady state Townsend and arrival time spectra experiments \citep{Dujko2008,Kondo1990}.
The skewness tensor can also be employed in fluid models of discharges,
by pairing a generalised diffusion equation, which includes the contributions
of the third order transport coefficients, with Poisson's equation.
This could be particularly important for discharges where ions play
an important role \citep{Petrovic2017}, or in situations where the
hydrodynamic approximation is at the limit of applicability, as in
the presence of sources and sinks of particles or in the close vicinity
of physical boundaries.

\begin{figure}
\includegraphics{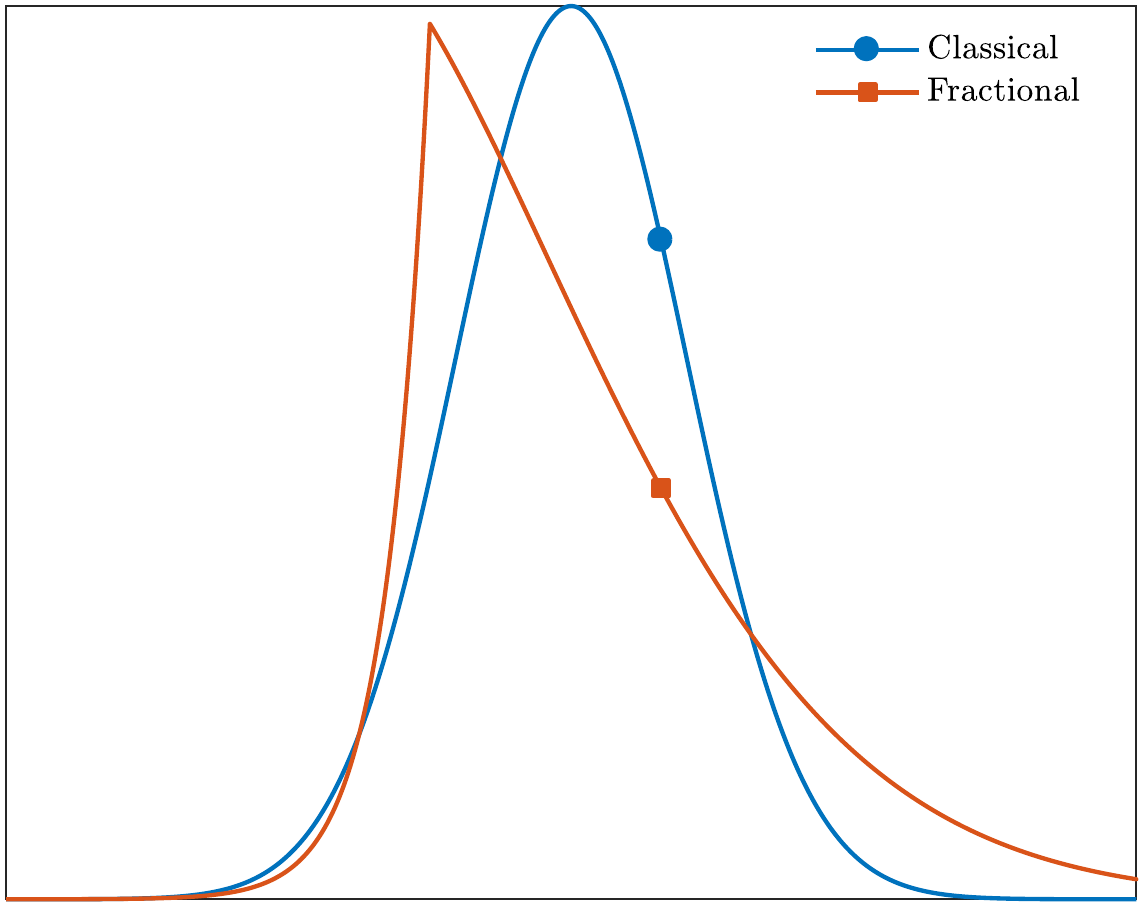}
\centering{}\caption{\label{fig:skewed}Skewed solution of the Caputo fractional advection-diffusion
equation alongside the corresponding Gaussian solution of the classical
advection-diffusion equation. Both pulses have evolved from an impulse
initial condition. The cusp in the fractional solution denotes the
location of this initial impulse.}
\end{figure}

In this manuscript, we are concerned with the form of the skewness
tensor for charged-particle transport in the presence of trapped (localised)
states. In particular, we are interested in the scenario where transport
is dispersive. Dispersive transport is characterised by a mean squared
displacement that increases sublinearly with time \citep{Metzler1999}.
Due to this non-integer power-law dependence, we refer to dispersive
transport as fractional transport throughout this manuscript. Some
examples of fractional transport include the trapping of charge carriers
in local imperfections in semiconductors \citep{Scher1975,Sibatov2007}
and both electron \citep{Mauracher2014,Borghesani2002,Sakai1992}
and positronium \citep{Stepanov2012,Stepanov2002,Charlton2001} trapping
in bubble states within liquids. Third order transport coefficients
are expected to be more sensitive to the influence of non-conservative
collisions than those of lower order, suggesting that the presence
of such trapped states would significantly influence the skewness
tensor. Indeed, Fig. \ref{fig:skewed} plots the solution of the Caputo
fractional advection-diffusion equation, a common model for fractional
transport \citep{Metzler1999}, and finds that it is skewed in comparison
to the Gaussian solution of the corresponding classical advection-diffusion
equation.

In Sec. \ref{sec:Model} of this study, we outline a general phase-space
kinetic model \citep{Stokes2016} for charged-particle transport via
localised and delocalised states. This model is capable of describing
both normal and fractional transport. We follow in Sec. \ref{sec:TransportCoefficients}
with a derivation of the flux transport coefficients for this model
up to third order. Sec. \ref{sec:Structure} explores the structure
of these transport coefficients and their symmetries under parity
transformation. The transport coefficients are used to extend Fick's
law, which leads to a generalised advection-diffusion-skewness equation,
presented in Sec. \ref{sec:DiffusionEquation}. In this section, we
also provide a physical interpretation of trap-induced skewness. By
analogy with Einstein's relation, Sec. \ref{sec:EinsteinRelation}
provides a relation between skewness, diffusion, mobility and temperature.
Sec. \ref{sec:FractionalTransport} looks at the case of fractional
transport and its effects on the flux transport coefficients. Finally,
Sec. \ref{sec:Conclusion} lists conclusions along with possible avenues
for future work.

\section{\label{sec:Model}Phase-space kinetic model}

We previously reported \citep{Stokes2016} the development of a phase-space
kinetic model wherein charged particles scatter due to collisions,
enter and leave traps and undergo recombination. In this model, free
particles are described by a phase-space distribution function $f\left(t,\mathbf{r},\mathbf{v}\right)$,
defined by the generalised Boltzmann equation
\begin{align}
\left(\frac{\partial}{\partial t}+\mathbf{v}\cdot\frac{\partial}{\partial\mathbf{r}}+\frac{e\mathbf{E}}{m}\cdot\frac{\partial}{\partial\mathbf{v}}\right)f\left(t,\mathbf{r},\mathbf{v}\right) & =-\nu_{\mathrm{coll}}\left[f\left(t,\mathbf{r},\mathbf{v}\right)-n\left(t,\mathbf{r}\right)w\left(T_{\mathrm{coll}},v\right)\right]\nonumber \\
 & -\nu_{\mathrm{trap}}\left[f\left(t,\mathbf{r},\mathbf{v}\right)-\Phi\left(t\right)\ast n\left(t,\mathbf{r}\right)w\left(T_{\mathrm{detrap}},v\right)\right]\nonumber \\
 & -\nu_{\mathrm{loss}}^{\left(\mathrm{free}\right)}f\left(t,\mathbf{r},\mathbf{v}\right),\label{eq:Boltzmann}
\end{align}
where $\mathbf{E}$ is the applied electric field and particles have
charge $e$, mass $m$ and number density $n\left(t,\mathbf{r}\right)\equiv\int\mathrm{d}\mathbf{v}f\left(t,\mathbf{r},\mathbf{v}\right)$.

Here, collisions, trapping and free particle recombination occur at
the constant frequencies $\nu_{\mathrm{coll}}$, $\nu_{\mathrm{trap}}$
and $\nu_{\mathrm{loss}}^{\left(\mathrm{free}\right)}$, respectively.
For collisions, the Bhatnagar\textendash Gross\textendash Krook (BGK)
collision operator \citep{Bhatnagar1954} has been used, which scatters
particles isotropically according to a Maxwellian velocity distribution
of temperature $T_{\mathrm{coll}}$. We define the Maxwellian velocity
distribution of temperature $T$ as
\begin{equation}
w\left(T,v\right)\equiv\left(\frac{m}{2\pi k_{\mathrm{B}}T}\right)^{\frac{3}{2}}\exp\left(-\frac{mv^{2}}{2k_{\mathrm{B}}T}\right),
\end{equation}
where $k_{\mathrm{B}}$ is the Boltzmann constant. Similarly, we use
the BGK-type operator introduced by Philippa \textit{et al}. \citep{Philippa2014}
to describe the processes of trapping and detrapping. This operator
specifies that particles leave traps with a Maxwellian distribution
of velocities of temperature $T_{\mathrm{detrap}}$ after a delay
that is governed by the distribution of trapping times $\phi\left(t\right)$.
This distribution appears in Eq. \eqref{eq:Boltzmann} through the
effective waiting time distribution
\begin{equation}
\Phi\left(t\right)\equiv\mathrm{e}^{-\nu_{\mathrm{loss}}^{\left(\mathrm{trap}\right)}t}\phi\left(t\right),
\end{equation}
that takes into account trapped particle recombination at the frequency
$\nu_{\mathrm{loss}}^{\left(\mathrm{trap}\right)}$.

\section{\label{sec:TransportCoefficients}Transport coefficients to third
order}

By integrating the generalised Boltzmann equation \eqref{eq:Boltzmann}
throughout all of velocity space, we find the equation of continuity
for the number density $n\left(t,\mathbf{r}\right)$:
\begin{equation}
\left[\frac{\partial}{\partial t}+\nu_{\mathrm{trap}}\left(1-\Phi\left(t\right)\ast\right)+\nu_{\mathrm{loss}}^{\left(\mathrm{free}\right)}\right]n\left(t,\mathbf{r}\right)+\frac{\partial}{\partial\mathbf{r}}\cdot\boldsymbol{\Gamma}\left(t,\mathbf{r}\right)=0,\label{eq:continuityEquation}
\end{equation}
where the particle flux is
\begin{equation}
\boldsymbol{\Gamma}\left(t,\mathbf{r}\right)\equiv\int\mathrm{d}\mathbf{v}\,\mathbf{v}f\left(t,\mathbf{r},\mathbf{v}\right).\label{eq:fluxDefinition}
\end{equation}
In the weak-gradient hydrodynamic regime, physical quantities can
be written as an infinite series of spatial gradients of the number
density $n\left(t,\mathbf{r}\right)$ \citep{Robson2006,Robson2017}.
In the case of the flux $\boldsymbol{\Gamma}\left(t,\mathbf{r}\right)$,
such a density gradient expansion provides a generalisation of Fick's
law:
\begin{equation}
\boldsymbol{\Gamma}=\mathbf{W}n-\boldsymbol{\mathsf{D}}\cdot\frac{\partial n}{\partial\mathbf{r}}+\boldsymbol{\mathsf{Q}}\colon\frac{\partial^{2}n}{\partial\mathbf{r}\partial\mathbf{r}}-\cdots,\label{eq:fluxExpansion}
\end{equation}
where $\mathbf{W}$ is the drift velocity vector, $\boldsymbol{\mathsf{D}}$
is the rank-2 diffusion tensor and $\boldsymbol{\mathsf{Q}}$ is the
rank-3 skewness tensor. To determine these flux transport coefficients
it is simply a matter of writing the solution of the generalised Boltzmann
equation \eqref{eq:Boltzmann} itself as a density gradient expansion
\begin{equation}
f\left(t,\mathbf{r},\mathbf{v}\right)=f^{\left(0\right)}\left(\mathbf{v}\right)n+\mathbf{f}^{\left(1\right)}\left(\mathbf{v}\right)\cdot\frac{\partial n}{\partial\mathbf{r}}+\boldsymbol{\mathsf{f}}^{\left(2\right)}\left(\mathbf{v}\right)\colon\frac{\partial^{2}n}{\partial\mathbf{r}\partial\mathbf{r}}+\cdots,\label{eq:fExpansion}
\end{equation}
and then evaluating the flux using Eq. \eqref{eq:fluxDefinition},
resulting in the transport coefficients
\begin{align}
\mathbf{W} & \equiv\int\mathrm{d}\mathbf{v}\,\mathbf{v}f^{\left(0\right)}\left(\mathbf{v}\right),\label{eq:driftIntegral}\\
\boldsymbol{\mathsf{D}} & \equiv\int\mathrm{d}\mathbf{v}\,\mathbf{v}\mathbf{f}^{\left(1\right)}\left(\mathbf{v}\right),\label{eq:diffusionIntegral}\\
\boldsymbol{\mathsf{Q}} & \equiv\int\mathrm{d}\mathbf{v}\,\mathbf{v}\boldsymbol{\mathsf{f}}^{\left(2\right)}\left(\mathbf{v}\right).\label{eq:skewnessIntegral}
\end{align}
Substituting the density gradient expansion of $f\left(t,\mathbf{r},\mathbf{v}\right)$
into the Boltzmann equation \eqref{eq:Boltzmann} and equating coefficients
of spatial gradients, as done in Sec. IV of Ref. \citep{Stokes2016},
gives the following coefficients
\begin{align}
f^{\left(0\right)}\left(\mathbf{s}\right) & =\frac{\nu_{\mathrm{coll}}w\left(\alpha_{\mathrm{coll}},s\right)+R\nu_{\mathrm{trap}}w\left(\alpha_{\mathrm{detrap}},s\right)}{\nu_{\mathrm{coll}}+R\nu_{\mathrm{trap}}+\frac{e\mathbf{E}}{m}\cdot\imath\mathbf{s}},\\
\mathbf{f}^{\left(1\right)}\left(\mathbf{s}\right) & =\frac{\nu_{\mathrm{trap}}\mathbf{R}^{\left(1\right)}w\left(\alpha_{\mathrm{detrap}},s\right)+f^{\left(0\right)}\left(\mathbf{s}\right)\left(\mathbf{W}-\nu_{\mathrm{trap}}\mathbf{R}^{\left(1\right)}\right)-\imath\frac{\partial f^{\left(0\right)}}{\partial\mathbf{s}}}{\nu_{\mathrm{coll}}+R\nu_{\mathrm{trap}}+\frac{e\mathbf{E}}{m}\cdot\imath\mathbf{s}},\\
\boldsymbol{\mathsf{f}}^{\left(2\right)}\left(\mathbf{s}\right) & =\frac{\nu_{\mathrm{trap}}\boldsymbol{\mathsf{R}}^{\left(2\right)}w\left(\alpha_{\mathrm{detrap}},s\right)-f^{\left(0\right)}\left(\mathbf{s}\right)\left(\boldsymbol{\mathsf{D}}+\nu_{\mathrm{trap}}\boldsymbol{\mathsf{R}}^{\left(2\right)}\right)+\mathbf{f}^{\left(1\right)}\left(\mathbf{s}\right)\left(\mathbf{W}-\nu_{\mathrm{trap}}\mathbf{R}^{\left(1\right)}\right)-\imath\frac{\partial\mathbf{f}^{\left(1\right)}}{\partial\mathbf{s}}}{\nu_{\mathrm{coll}}+R\nu_{\mathrm{trap}}+\frac{e\mathbf{E}}{m}\cdot\imath\mathbf{s}},
\end{align}
where a Fourier transform has been performed in velocity space, $f\left(\mathbf{s}\right)\equiv\int\mathrm{d}\mathbf{v}\mathrm{e}^{-\imath\mathbf{s}\cdot\mathbf{v}}f\left(\mathbf{v}\right)$.
As in Ref. \citep{Stokes2016}, we have used the density gradient
expansion of the concentration of particles leaving traps:
\begin{equation}
\Phi\left(t\right)\ast n\left(t,\mathbf{r}\right)=Rn+\mathbf{R}^{\left(1\right)}\cdot\frac{\partial n}{\partial\mathbf{r}}+\boldsymbol{\mathsf{R}}^{\left(2\right)}\colon\frac{\partial^{2}n}{\partial\mathbf{r}\partial\mathbf{r}}+\cdots,\label{eq:detrapConcentration}
\end{equation}
the coefficients of which are related to the flux transport coefficients
through
\begin{align}
\mathbf{R}^{\left(1\right)} & \equiv\frac{R\left\langle t\right\rangle }{1+\nu_{\mathrm{trap}}R\left\langle t\right\rangle }\mathbf{W},\\
\boldsymbol{\mathsf{R}}^{\left(2\right)} & \equiv\frac{R\left\langle t^{2}\right\rangle }{2\left(1+\nu_{\mathrm{trap}}R\left\langle t\right\rangle \right)^{3}}\mathbf{W}\mathbf{W}-\frac{R\left\langle t\right\rangle }{1+\nu_{\mathrm{trap}}R\left\langle t\right\rangle }\boldsymbol{\mathsf{D}},
\end{align}
where the time averages are defined
\begin{equation}
\left\langle \eta\left(t\right)\right\rangle \equiv\frac{1}{R}\int_{0}^{\infty}\mathrm{d}t\Phi\left(t\right)\mathrm{e}^{\left[\nu_{\mathrm{loss}}^{\left(\mathrm{free}\right)}+\nu_{\mathrm{trap}}\left(1-R\right)\right]t}\eta\left(t\right).\label{eq:timeAverage}
\end{equation}
Applying this time average to unity results in an implicit definition
for the initial coefficient $R$: 
\begin{equation}
R\equiv\int_{0}^{\infty}\mathrm{d}t\Phi\left(t\right)\mathrm{e}^{\left[\nu_{\mathrm{loss}}^{\left(\mathrm{free}\right)}+\nu_{\mathrm{trap}}\left(1-R\right)\right]t}.
\end{equation}
Thus, for every trapping time distribution $\phi\left(t\right)$ there
corresponds a value of $R$. Some values are tabulated in Appendix
A of Ref. \citep{Stokes2016}.

Proceeding to evaluate Eqs. \eqref{eq:driftIntegral}\textendash \eqref{eq:skewnessIntegral}
for the transport coefficients, we find
\begin{align}
\mathbf{W} & \equiv\frac{e\mathbf{E}}{m\nu_{\mathrm{eff}}},\label{eq:drift}\\
\boldsymbol{\mathsf{D}} & \equiv\frac{1}{\nu_{\mathrm{eff}}}\left(\frac{k_{\mathrm{B}}T_{\mathrm{eff}}}{m}\boldsymbol{\mathsf{I}}+\frac{1+2\nu_{\mathrm{trap}}R\left\langle t\right\rangle }{1+\nu_{\mathrm{trap}}R\left\langle t\right\rangle }\mathbf{W}\mathbf{W}\right),\label{eq:diffusion}\\
\boldsymbol{\mathsf{Q}} & \equiv\left[1+\left(\frac{1+2\nu_{\mathrm{trap}}R\left\langle t\right\rangle }{1+\nu_{\mathrm{trap}}R\left\langle t\right\rangle }\right)^{2}-\frac{\nu_{\mathrm{trap}}R\left\langle t^{2}\right\rangle }{4\left(1+\nu_{\mathrm{trap}}R\left\langle t\right\rangle \right)^{3}}\nu_{\mathrm{eff}}\right]\frac{2\mathbf{W}\mathbf{W}\mathbf{W}}{\nu_{\mathrm{eff}}^{2}}\nonumber \\
 & +\frac{1+2\nu_{\mathrm{trap}}R\left\langle t\right\rangle }{1+\nu_{\mathrm{trap}}R\left\langle t\right\rangle }\frac{k_{\mathrm{B}}T_{\mathrm{eff}}}{m\nu_{\mathrm{eff}}^{2}}\left(\boldsymbol{\mathsf{I}}\mathbf{W}+\mathbf{e}_{1}\mathbf{W}\mathbf{e}_{1}+\mathbf{e}_{2}\mathbf{W}\mathbf{e}_{2}+\mathbf{e}_{3}\mathbf{W}\mathbf{e}_{3}\right)\nonumber \\
 & +\frac{\nu_{\mathrm{trap}}R\left\langle t\right\rangle }{1+\nu_{\mathrm{trap}}R\left\langle t\right\rangle }\frac{\nu_{\mathrm{coll}}}{\nu_{\mathrm{eff}}}\frac{k_{\mathrm{B}}\left(T_{\mathrm{coll}}-T_{\mathrm{detrap}}\right)}{m\nu_{\mathrm{eff}}}\frac{\mathbf{W}\boldsymbol{\mathsf{I}}}{\nu_{\mathrm{eff}}},\label{eq:skewness}
\end{align}
where $\mathbf{e}_{1}$, $\mathbf{e}_{2}$ and $\mathbf{e}_{3}$ are
standard orthonormal basis vectors and we have introduced the effective
frequency and temperature:
\begin{align}
\nu_{\mathrm{eff}} & \equiv\nu_{\mathrm{coll}}+R\nu_{\mathrm{trap}},\\
T_{\mathrm{eff}} & \equiv\frac{\nu_{\mathrm{coll}}T_{\mathrm{coll}}+R\nu_{\mathrm{trap}}T_{\mathrm{detrap}}}{\nu_{\mathrm{coll}}+R\nu_{\mathrm{trap}}}.
\end{align}
We confirm that when there are no traps present, $\nu_{\mathrm{trap}}=0$,
the transport coefficients agree with those of the BGK collision model,
previously found by Robson \citep{Robson1975}:
\begin{align}
\mathbf{W} & \equiv\frac{e\mathbf{E}}{m\nu_{\mathrm{coll}}},\\
\boldsymbol{\mathsf{D}} & \equiv\frac{1}{\nu_{\mathrm{coll}}}\left(\frac{k_{\mathrm{B}}T_{\mathrm{coll}}}{m}\boldsymbol{\mathsf{I}}+\mathbf{W}\mathbf{W}\right),\\
\boldsymbol{\mathsf{Q}} & \equiv\frac{1}{\nu_{\mathrm{coll}}^{2}}\left[\frac{k_{\mathrm{B}}T_{\mathrm{coll}}}{m}\left(\boldsymbol{\mathsf{I}}\mathbf{W}+\mathbf{e}_{1}\mathbf{W}\mathbf{e}_{1}+\mathbf{e}_{2}\mathbf{W}\mathbf{e}_{2}+\mathbf{e}_{3}\mathbf{W}\mathbf{e}_{3}\right)+4\mathbf{W}\mathbf{W}\mathbf{W}\right].\label{eq:BGKskewness}
\end{align}

\section{\label{sec:Structure}Structure and symmetry of transport coefficients}

If we align the basis vector $\mathbf{e}_{3}$ parallel to the applied
electric field $\mathbf{E}$, the transport coefficients \eqref{eq:drift}\textendash \eqref{eq:skewness}
take on the known tensor structure \citep{Robson2006,Robson2017,Whealton1974,Vrhovac1999,Koutselos2001}:
\begin{align}
\mathbf{W} & \equiv\left[\begin{array}{c}
0\\
0\\
W
\end{array}\right],\label{eq:W}\\
\boldsymbol{\mathsf{D}} & \equiv\left[\begin{array}{ccc}
D_{\perp} & 0 & 0\\
0 & D_{\perp} & 0\\
0 & 0 & D_{\parallel}
\end{array}\right]\\
\boldsymbol{\mathsf{Q}}_{xab} & \equiv\left[\begin{array}{ccc}
0 & 0 & Q_{1}\\
0 & 0 & 0\\
Q_{1} & 0 & 0
\end{array}\right],\label{eq:xSkewness}\\
\boldsymbol{\mathsf{Q}}_{yab} & \equiv\left[\begin{array}{ccc}
0 & 0 & 0\\
0 & 0 & Q_{1}\\
0 & Q_{1} & 0
\end{array}\right],\label{eq:ySkewness}\\
\boldsymbol{\mathsf{Q}}_{zab} & \equiv\left[\begin{array}{ccc}
Q_{2} & 0 & 0\\
0 & Q_{2} & 0\\
0 & 0 & 2Q_{1}+Q_{2}+Q_{3}
\end{array}\right],\label{eq:zSkewness}
\end{align}
where $a,b\in\left\{ x,y,z\right\} $. Here, the drift velocity is
defined by the speed
\begin{equation}
W\equiv\frac{eE}{m\nu_{\mathrm{eff}}},
\end{equation}
the diffusion coefficient is defined by two components perpendicular
and parallel to the field
\begin{align}
D_{\perp} & \equiv\frac{k_{\mathrm{B}}T_{\mathrm{eff}}}{m\nu_{\mathrm{eff}}},\\
D_{\parallel} & \equiv D_{\perp}+\frac{1+2\nu_{\mathrm{trap}}R\left\langle t\right\rangle }{1+\nu_{\mathrm{trap}}R\left\langle t\right\rangle }\frac{W^{2}}{\nu_{\mathrm{eff}}},\label{eq:diffusionPara}
\end{align}
and the skewness is defined by the three independent components
\begin{align}
Q_{1} & \equiv\frac{1+2\nu_{\mathrm{trap}}R\left\langle t\right\rangle }{1+\nu_{\mathrm{trap}}R\left\langle t\right\rangle }\frac{k_{\mathrm{B}}T_{\mathrm{eff}}}{m\nu_{\mathrm{eff}}}\frac{W}{\nu_{\mathrm{eff}}},\label{eq:component1}\\
Q_{2} & \equiv\frac{\nu_{\mathrm{trap}}R\left\langle t\right\rangle }{1+\nu_{\mathrm{trap}}R\left\langle t\right\rangle }\frac{\nu_{\mathrm{coll}}}{\nu_{\mathrm{eff}}}\frac{k_{\mathrm{B}}\left(T_{\mathrm{coll}}-T_{\mathrm{detrap}}\right)}{m\nu_{\mathrm{eff}}}\frac{W}{\nu_{\mathrm{eff}}},\label{eq:component2}\\
Q_{3} & \equiv\left[1+\left(\frac{1+2\nu_{\mathrm{trap}}R\left\langle t\right\rangle }{1+\nu_{\mathrm{trap}}R\left\langle t\right\rangle }\right)^{2}-\frac{\nu_{\mathrm{trap}}R\left\langle t^{2}\right\rangle }{4\left(1+\nu_{\mathrm{trap}}R\left\langle t\right\rangle \right)^{3}}\nu_{\mathrm{eff}}\right]\frac{2W^{3}}{\nu_{\mathrm{eff}}^{2}}.\label{eq:component3}
\end{align}
Although this is the case in general, there are situations where the
skewness can be defined using fewer than three components. Indeed,
this is the case for the BGK model as studied by Robson \citep{Robson1975}
where the skewness given by Eq. \eqref{eq:BGKskewness} is defined
using only the components $Q_{1}$ and $Q_{3}$, with $Q_{2}=0$.
The component $Q_{2}$ vanishes in this case due to the simple Maxwellian
source term used to describe scattered particles. For $Q_{2}$ to
arise, it is necessary that this source term has some spatial dependence,
as occurs for our model through the concentration of particles leaving
traps, $\Phi\left(t\right)\ast n\left(t,\mathbf{r}\right)$, and its
density gradient expansion \eqref{eq:detrapConcentration}.

Lastly, we also confirm that the symmetry of transport coefficients
with respect to the parity transformation $\mathbf{E}\rightarrow-\mathbf{E}$
depends on the parity of the order of each transport coefficient \citep{Whealton1974,White1999}:
\begin{align}
\mathbf{W} & \rightarrow-\mathbf{W},\\
\boldsymbol{\mathsf{D}} & \rightarrow\boldsymbol{\mathsf{D}},\\
\boldsymbol{\mathsf{Q}} & \rightarrow-\boldsymbol{\mathsf{Q}}.\label{eq:skewnessParity}
\end{align}

\section{\label{sec:DiffusionEquation}Generalised advection-diffusion-skewness
equation}

Using the density gradient expansion \eqref{eq:fluxExpansion} for
the flux $\boldsymbol{\Gamma}\left(t,\mathbf{r}\right)$ up to second
spatial order in conjunction with the continuity equation \eqref{eq:continuityEquation}
results in the generalised advection-diffusion-skewness equation
\begin{equation}
\left[\frac{\partial}{\partial t}+\nu_{\mathrm{trap}}\left(1-\Phi\left(t\right)\ast\right)+\nu_{\mathrm{loss}}^{\left(\mathrm{free}\right)}\right]n\left(t,\mathbf{r}\right)+\mathbf{W}\cdot\frac{\partial n}{\partial\mathbf{r}}-\boldsymbol{\mathsf{D}}\colon\frac{\partial^{2}n}{\partial\mathbf{r}\partial\mathbf{r}}+\boldsymbol{\mathsf{Q}}\vdots\frac{\partial^{3}n}{\partial\mathbf{r}\partial\mathbf{r}\partial\mathbf{r}}=0,\label{eq:ADSequation}
\end{equation}
valid in the weak-gradient hydrodynamic regime. In Cartesian coordinates
$\left(x,y,z\right)$ with the electric field $\mathbf{E}$ aligned
in the $z$-direction, the transport coefficients take the form of
Eqs. \eqref{eq:W}\textendash \eqref{eq:zSkewness} and the advection-diffusion-skewness
equation becomes
\begin{align}
\left[\frac{\partial}{\partial t}+\nu_{\mathrm{trap}}\left(1-\Phi\left(t\right)\ast\right)+\nu_{\mathrm{loss}}^{\left(\mathrm{free}\right)}\right]n\left(t,x,y,z\right)+W\frac{\partial n}{\partial z}-D_{\perp}\left(\frac{\partial^{2}n}{\partial x^{2}}+\frac{\partial^{2}n}{\partial y^{2}}\right)-D_{\parallel}\frac{\partial^{2}n}{\partial z^{2}}\nonumber \\
+3Q_{\perp}\left(\frac{\partial^{2}}{\partial x^{2}}+\frac{\partial^{2}}{\partial y^{2}}\right)\frac{\partial n}{\partial z}+Q_{\parallel}\frac{\partial^{3}n}{\partial z^{3}} & =0,\label{eq:ADScartesian}
\end{align}
where the skewness manifests as components perpendicular and parallel
to the applied field \citep{Petrovic2017,Koutselos2001,Vrhovac1999}:
\begin{align}
Q_{\perp} & \equiv\frac{Q_{zxx}+Q_{xzx}+Q_{xxz}}{3},\\
Q_{\parallel} & \equiv Q_{zzz},
\end{align}
which in terms of the independent components \eqref{eq:component1}\textendash \eqref{eq:component3}
are
\begin{align}
Q_{\perp} & =\frac{2Q_{1}+Q_{2}}{3},\\
Q_{\parallel} & =2Q_{1}+Q_{2}+Q_{3}.
\end{align}
Written in full, the perpendicular and parallel skewnesses are
\begin{align}
Q_{\perp} & =\frac{2D_{\perp}W}{3\nu_{\mathrm{eff}}}\nonumber \\
 & +\frac{\nu_{\mathrm{trap}}R\left\langle t\right\rangle }{1+\nu_{\mathrm{trap}}R\left\langle t\right\rangle }\left(D_{\perp}-\frac{k_{\mathrm{B}}T_{\mathrm{detrap}}}{3m\nu_{\mathrm{eff}}}\right)\frac{W}{\nu_{\mathrm{eff}}},\label{eq:skewnessPerp}\\
Q_{\parallel} & =3Q_{\perp}+\frac{4W^{3}}{\nu_{\mathrm{eff}}^{2}}\nonumber \\
 & +\frac{\nu_{\mathrm{trap}}R\left\langle t\right\rangle }{1+\nu_{\mathrm{trap}}R\left\langle t\right\rangle }\left[6-\frac{2}{1+\nu_{\mathrm{trap}}R\left\langle t\right\rangle }-\frac{\nu_{\mathrm{eff}}\left\langle t^{2}\right\rangle }{2\left\langle t\right\rangle \left(1+\nu_{\mathrm{trap}}R\left\langle t\right\rangle \right)^{2}}\right]\frac{W^{3}}{\nu_{\mathrm{eff}}^{2}},\label{eq:skewnessPara}
\end{align}
where terms present due to trapping have been grouped separately.
An alternative form of the skewness tensor that makes use of these
components explicitly is
\begin{align}
\tilde{\boldsymbol{\mathsf{Q}}}_{xab} & \equiv\left[\begin{array}{ccc}
0 & 0 & 0\\
0 & 0 & 0\\
0 & 0 & 0
\end{array}\right],\\
\tilde{\boldsymbol{\mathsf{Q}}}_{yab} & \equiv\left[\begin{array}{ccc}
0 & 0 & 0\\
0 & 0 & 0\\
0 & 0 & 0
\end{array}\right],\\
\tilde{\boldsymbol{\mathsf{Q}}}_{zab} & \equiv\left[\begin{array}{ccc}
3Q_{\perp} & 0 & 0\\
0 & 3Q_{\perp} & 0\\
0 & 0 & Q_{\parallel}
\end{array}\right],
\end{align}
where $a,b\in\left\{ x,y,z\right\} $. This form was used by Robson
\citep{Robson1975} when expressing the BGK model skewness \eqref{eq:BGKskewness}
and is valid only when the skewness is triple-contracted with a symmetric
tensor, as occurs in the advection-diffusion-skewness equation \eqref{eq:ADSequation}.

\begin{figure}
\includegraphics[scale=0.25]{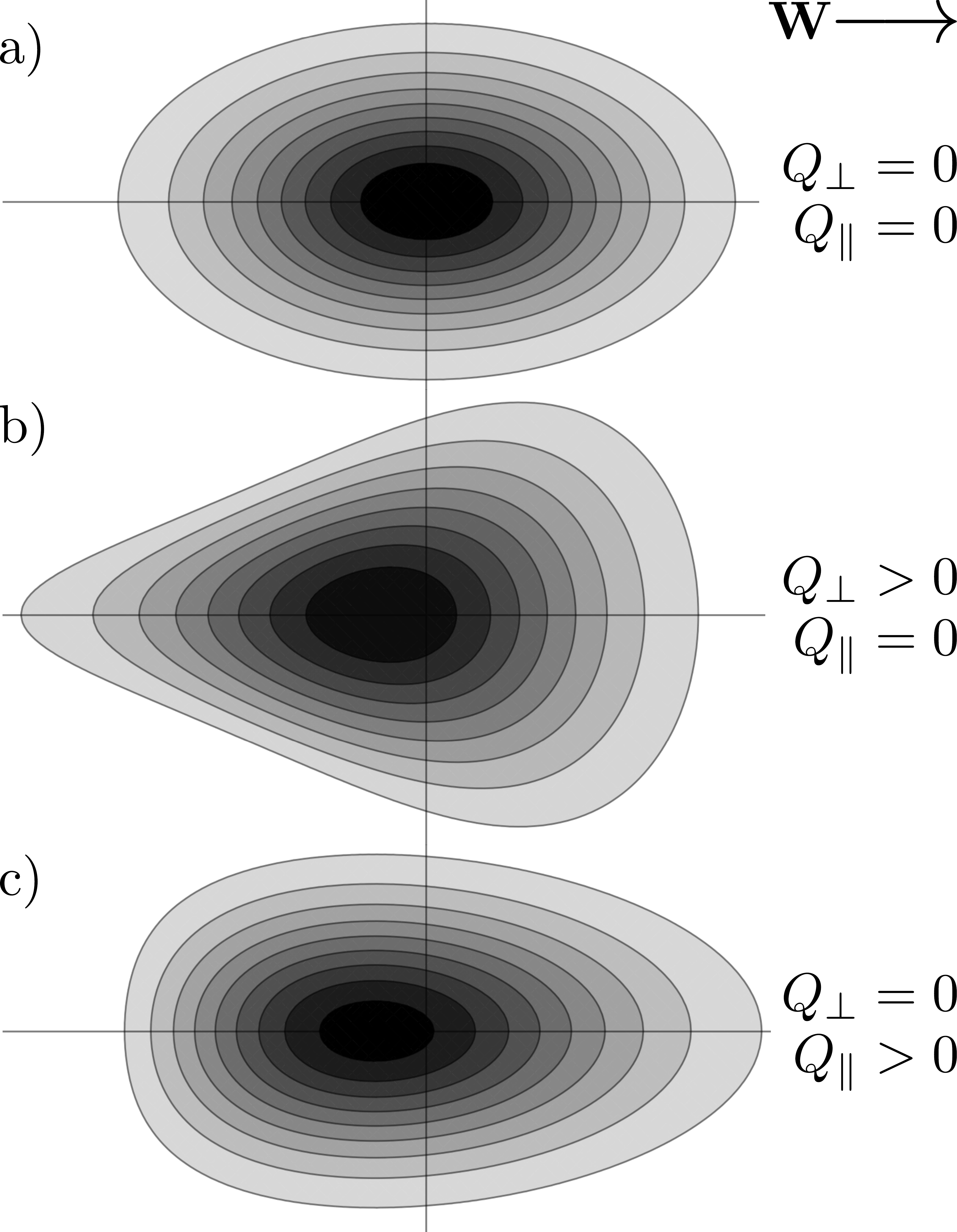}
\centering{}\caption{\label{fig:skewness2D}Contours of constant number density as defined
by the advection-diffusion-skewness equation \eqref{eq:ADScartesian}
with drift velocity $\mathbf{W}$ and anisotropic diffusion $D_{\parallel}>D_{\perp}>0$
for no skewness, a), positive perpendicular skewness, b), and positive
parallel skewness, c). Each profile has evolved from an impulse initial
condition. As the skewness tensor is odd under parity transformation,
Eq. \eqref{eq:skewnessParity}, the case of negative skewness can
be considered by reflecting the above profiles horizontally across
the vertical axis.}
\end{figure}

To provide some physical intuition regarding the perpendicular and
parallel skewness coefficients, $Q_{\perp}$ and $Q_{\parallel}$,
we solve the advection-diffusion-skewness equation \eqref{eq:ADScartesian}
for an impulse initial condition and perform contour plots of the
resulting pulse in Fig. \ref{fig:skewness2D}. Fig. \ref{fig:skewness2D}
a) considers the case of no skewness, $Q_{\perp}=Q_{\parallel}=0$,
and displays the expected Gaussian solution with elliptical contours
due to anisotropic diffusion. Fig. \ref{fig:skewness2D} b) and c)
consider positive perpendicular and parallel skewnesses, respectively.
In both cases, it can be seen that skewness introduces asymmetry in
the pulse in the direction of the field. In general, positive skewness
can be seen to reduce the spread of particles behind the pulse, while
enhancing the spread toward the front of the pulse. In Fig. \ref{fig:skewness2D}
b) for positive perpendicular skewness, this change in particle spread
primarily occurs transverse to the field, resulting in a vaguely triangular
pulse profile. In Fig. \ref{fig:skewness2D} c) for positive parallel
skewness, this change in particle spread occurs longitudinally which,
in the language of statistics, results in a distribution with positive
skew.

\begin{figure}
\includegraphics{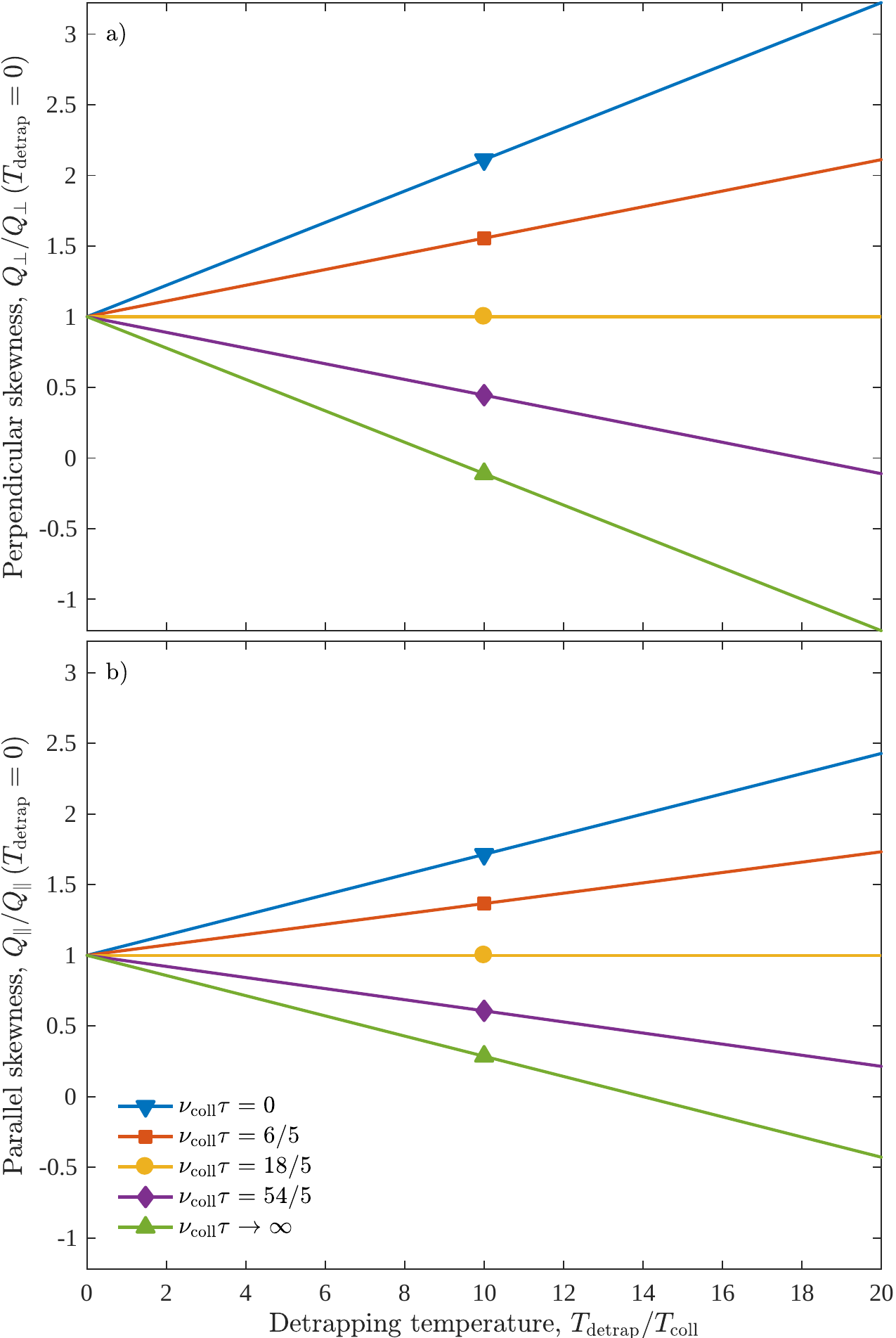}
\centering{}\caption{\label{fig:skewnessPerpPara}Linear plots of perpendicular and parallel
skewness coefficients, $Q_{\perp}$ and $Q_{\parallel}$, versus the
detrapping temperature $T_{\mathrm{detrap}}$. Here, traps are described
by an exponential distribution of trapping times $\phi\left(t\right)=\frac{1}{\tau}\exp\left(-\frac{t}{\tau}\right)$,
and no recombination is considered, $\nu_{\mathrm{loss}}^{\left(\mathrm{free}\right)}=\nu_{\mathrm{loss}}^{\left(\mathrm{trap}\right)}=0$.
To perform these plots, we choose a trapping frequency of $\nu_{\mathrm{trap}}/\nu_{\mathrm{coll}}=1/9$,
while b) also requires that we specify a drift velocity $\mathbf{W}$,
which we choose such that $mW^{2}/k_{\mathrm{B}}T_{\mathrm{coll}}=1/4$.
The gradients in b) are of smaller magnitude than a) due to the greater
dependence of the parallel skewness \eqref{eq:skewnessPara} on the
drift speed $W$ as compared to the perpendicular skewness \eqref{eq:skewnessPerp}.
Thus, as the drift speed decreases, the plots in b) coincide with
those in a). When detrapping is instantaneous, $\tau=0$, the skewness
gradients are positive, implying that the skewness caused by traps
is also positive. As the mean trapping time $\tau$ increases, the
skewness gradients decrease, becoming negative and implying a corresponding
negative skewness due to traps. The limiting case of an infinite mean
trapping time, $\tau\rightarrow\infty$, corresponds to fractional
transport, which is the subject of Sec. \ref{sec:FractionalTransport}.
We observe from this figure that the skewness coefficients become
overall negative when particles leave traps with a sufficiently large
temperature $T_{\mathrm{detrap}}$ after a sufficiently long amount
of time $\tau$. This observation coincides with the illustration
of skewness in Fig. \ref{fig:skewness2D} where negative skewness
is characterised by an increased particle spread behind the pulse,
which we attribute here to particles returning from traps.}
\end{figure}

In our previous manuscript \citep{Stokes2016}, we interpreted the
trap-induced anisotropic diffusion present in Eq. \eqref{eq:diffusionPara}
as a consequence of the physical separation between trapped particles
and free particles moving with the field. In a similar fashion, we
can interpret the trap-induced skewness present in the perpendicular
and parallel skewness coefficients \eqref{eq:skewnessPerp} and \eqref{eq:skewnessPara}.
To achieve this, we plot the skewness against the detrapping temperature
$T_{\mathrm{detrap}}$ for various mean trapping times in Fig. \ref{fig:skewnessPerpPara}.
The resulting plots are linear with gradients that characterise of
the type of skewness caused by traps. That is, positive or negative
gradients correspond respectively to positive or negative trap-based
skewness.

When the mean trapping time is zero, the gradients in Fig. \ref{fig:skewnessPerpPara}
are positive and traps cause positive skewness. This is to be expected
as, in this case, trapping and detrapping simply act as an elastic
scattering process with a positive skewness akin to Eq. \eqref{eq:BGKskewness}
for the BGK collision model. As the mean trapping time increases,
the nature of the skewness caused by traps changes, ultimately becoming
negative for the parameters considered in Fig. \ref{fig:skewnessPerpPara}.
As illustrated in Fig. \ref{fig:skewness2D}, negative skewness corresponds
to an increased spread of particles behind the pulse. We interpret
the increased spread here as being due to particles returning from
traps. This interpretation implies that the skewness coefficients
could become overall negative if particles remain trapped for a sufficient
length of time before returning with a sufficiently large temperature.
Indeed, these are the conditions for which the skewness coefficients
become negative in Fig. \ref{fig:skewnessPerpPara}.

This phenomenon of negative skewness has been observed previously
by Petrovi\'{c} \textit{et al}. \citep{Petrovic2017} in the calculation
of the perpendicular skewness of electrons in methane. Only collisions
were considered in this study and so trapping is evidently not a necessary
condition for negative skewness to occur. However, it should be emphasised
that the skewness is strictly positive when collisions are described
by the simple BGK collision operator, as is seen in Eq. \eqref{eq:BGKskewness}.

\section{\label{sec:EinsteinRelation}Relating skewness, mobility and temperature}

The classical Einstein relation between diffusion, mobility and temperature
is \citep{Einstein1905}
\begin{equation}
\frac{\boldsymbol{\mathsf{D}}}{K}=\frac{k_{\mathrm{B}}\boldsymbol{\mathsf{T}}}{e},
\end{equation}
where $K$ is the mobility defined as satisfying $\mathbf{W}\equiv K\mathbf{E}$
and $\boldsymbol{\mathsf{T}}$ is the rank-2 temperature tensor. As
seen by Eq. \eqref{eq:diffusion} for the diffusion coefficient, the
phase-space model described by Eq. \eqref{eq:Boltzmann} has an enhanced
diffusivity in the direction of the field due to trapping and detrapping.
This enhancement manifests as the following generalised Einstein relation
\citep{Stokes2016a}
\begin{equation}
\frac{\boldsymbol{\mathsf{D}}}{K}=\frac{k_{\mathrm{B}}\boldsymbol{\mathsf{T}}}{e}+\frac{\nu_{\mathrm{trap}}R\left\langle t\right\rangle }{1+\nu_{\mathrm{trap}}R\left\langle t\right\rangle }\frac{m\mathbf{W}\mathbf{W}}{e}.
\end{equation}
By relating the skewness to the temperature tensor though this diffusion
coefficient, we find a skewness analogue to the Einstein relation:
\begin{align}
\boldsymbol{\mathsf{Q}} & \equiv\left[1-\frac{\nu_{\mathrm{trap}}R\left\langle t^{2}\right\rangle }{4\left(1+\nu_{\mathrm{trap}}R\left\langle t\right\rangle \right)^{3}}\nu_{\mathrm{eff}}\right]\frac{2\mathbf{W}\mathbf{W}\mathbf{W}}{\nu_{\mathrm{eff}}^{2}}\nonumber \\
 & +\frac{1+2\nu_{\mathrm{trap}}R\left\langle t\right\rangle }{1+\nu_{\mathrm{trap}}R\left\langle t\right\rangle }\frac{\boldsymbol{\mathsf{D}}\mathbf{W}+D_{\perp}\mathbf{e}_{1}\mathbf{W}\mathbf{e}_{1}+D_{\perp}\mathbf{e}_{2}\mathbf{W}\mathbf{e}_{2}+D_{\parallel}\mathbf{e}_{3}\mathbf{W}\mathbf{e}_{3}}{\nu_{\mathrm{eff}}}\nonumber \\
 & +\frac{\nu_{\mathrm{trap}}R\left\langle t\right\rangle }{1+\nu_{\mathrm{trap}}R\left\langle t\right\rangle }\frac{\nu_{\mathrm{coll}}}{\nu_{\mathrm{eff}}}\frac{k_{\mathrm{B}}\left(T_{\mathrm{coll}}-T_{\mathrm{detrap}}\right)}{m\nu_{\mathrm{eff}}}\frac{\mathbf{W}\boldsymbol{\mathsf{I}}}{\nu_{\mathrm{eff}}}.\label{eq:generalisedEinstein}
\end{align}

\section{\label{sec:FractionalTransport}The case of fractional transport}

For the phase-space kinetic model described by Eq. \eqref{eq:Boltzmann},
fractional transport can occur when the distribution of trapping times
has a heavy power-law tail of the form \citep{Stokes2016}
\begin{equation}
\phi\left(t\right)\sim t^{-\left(1+\alpha\right)}.\label{eq:heavyTail}
\end{equation}
Note that, as transport here is dispersive in nature, the mean trapping
time diverges:
\begin{equation}
\int_{0}^{\infty}\mathrm{d}t\phi\left(t\right)t\rightarrow\infty.
\end{equation}
Consequently, the time averages defined by Eq. \eqref{eq:timeAverage}
also diverge, correspondingly affecting the transport coefficients.
Thus, for fractional transport, the transport coefficients \eqref{eq:drift}\textendash \eqref{eq:skewness}
take on the simpler form \citep{Stokes2016}
\begin{align}
\mathbf{W} & =\frac{e\mathbf{E}}{m\nu_{\mathrm{eff}}},\label{eq:fractionalDrift}\\
\boldsymbol{\mathsf{D}} & =\frac{1}{\nu_{\mathrm{eff}}}\left(\frac{k_{\mathrm{B}}T_{\mathrm{eff}}}{m}\boldsymbol{\mathsf{I}}+2\mathbf{W}\mathbf{W}\right),\label{eq:fractionalDiffusion}\\
\boldsymbol{\mathsf{Q}} & =\frac{2\mathbf{W}\mathbf{W}\mathbf{W}}{\nu_{\mathrm{eff}}^{2}}\nonumber \\
 & +\frac{2\left(\boldsymbol{\mathsf{D}}\mathbf{W}+D_{\perp}\mathbf{e}_{1}\mathbf{W}\mathbf{e}_{1}+D_{\perp}\mathbf{e}_{2}\mathbf{W}\mathbf{e}_{2}+D_{\parallel}\mathbf{e}_{3}\mathbf{W}\mathbf{e}_{3}\right)}{\nu_{\mathrm{eff}}}\nonumber \\
 & +\frac{\nu_{\mathrm{coll}}}{\nu_{\mathrm{eff}}}\frac{k_{\mathrm{B}}\left(T_{\mathrm{coll}}-T_{\mathrm{detrap}}\right)}{m\nu_{\mathrm{eff}}}\frac{\mathbf{W}\boldsymbol{\mathsf{I}}}{\nu_{\mathrm{eff}}},\label{eq:fractionalSkewness}
\end{align}
where the effective frequency is now defined
\begin{equation}
\nu_{\mathrm{eff}}\equiv\nu_{\mathrm{coll}}+\nu_{\mathrm{trap}}+\nu_{\mathrm{loss}}^{\left(\mathrm{free}\right)}.
\end{equation}
Note that transport coefficients are now independent of the specific
choice of the trapping time distribution $\phi\left(t\right)$, so
long as the condition \eqref{eq:heavyTail} for fractional transport
is satisfied.

\section{\label{sec:Conclusion}Conclusion}

We have explored the transport coefficients of a phase-space kinetic
model \eqref{eq:Boltzmann} for both localised and delocalised transport.
In particular, we have considered up to the third-order transport
coefficient of skewness $\boldsymbol{\mathsf{Q}}$, which takes the
form of a rank-3 tensor. The structure of the skewness tensor and
its symmetry under parity transformation was found to be in agreement
with previous studies. These transport coefficients provide an extension
to Fick's law, Eq. \eqref{eq:fluxExpansion}, which we used to form
a generalised advection-diffusion-skewness equation \eqref{eq:ADSequation}
with a non-local time operator. We observed trap-induced negative
skewness and provided a corresponding physical interpretation. In
addition, by analogy with Einstein's relation, the skewness was related
to the mobility and temperature through Eq. \eqref{eq:generalisedEinstein}.
Lastly, the form of the transport coefficients for the particular
case of fractional transport were outlined in Eqs. \eqref{eq:fractionalDrift}\textendash \eqref{eq:fractionalSkewness}.

There exist a number of possibilities for future work. The focus of
this manuscript was on constant transport coefficients that define
the flux in the hydrodynamic regime as the density gradient expansion
\eqref{eq:fluxExpansion}. Transient transport coefficients and transport
coefficients of the bulk were not considered. Ref. \citep{Stokes2016}
outlines an analytical solution of the kinetic model \eqref{eq:Boltzmann}
that could be used to compute such transport coefficients through
time-varying velocity and spatial moments of the phase-space distribution
function $f\left(t,\mathbf{r},\mathbf{v}\right)$.

Another extension to this work could be to explore what consequences
energy-dependent collision, trapping and recombination frequencies
have on the skewness. Such a generalisation for Eq. \eqref{eq:Boltzmann}
was the focus of Ref. \citep{Stokes2016a}. This would allow for the
derivation of a skewness analogue of Einstein's relation that would
also take into account the field dependence of mobility \citep{Stokes2016a}.
This may also shed light on the recent results of Petrovi\'{c} \textit{et
al}. \citep{Petrovic2017}, that suggest a correlation between the
energy-dependent phenomenon of negative differential conductivity
and skewness.

Lastly, it is important to note that the extension to Fick's law described
in this paper is only useful when an electric field is present. Without
an applied field, the drift velocity, skewness and all other odd-ordered
transport coefficients would vanish. If we wish to extend Fick's law
in such a situation, we must also consider the kurtosis coefficient,
the next even-ordered transport coefficient beyond diffusion. The
kurtosis can be found in a straightforward fashion from the rank-3
tensorial coefficient $\boldsymbol{\mathsf{f}}^{\left(3\right)}\left(\mathbf{v}\right)$
in the density gradient expansion \eqref{eq:fExpansion} of the phase-space
distribution function $f\left(t,\mathbf{r},\mathbf{v}\right)$, in
the same way drift velocity, diffusion and skewness were found using
Eqs. \eqref{eq:driftIntegral}\textendash \eqref{eq:skewnessIntegral}.
\begin{acknowledgments}

\appendix
The authors gratefully acknowledge the useful discussions with Prof.
Robert Robson and the financial support of the Australian Research
Council.

IS and SD are supported by the Grants No. ON171037 and III41011 from
the Ministry of Education, Science and Technological Development of
the Republic of Serbia.

PS is supported by an Australian Government Research Training Program
Scholarship.
\end{acknowledgments}

\bibliography{references}

\end{document}